\documentclass[prd,aps,superscriptaddress]{revtex4}

\usepackage{graphicx}
\usepackage{dcolumn}
\usepackage{bm}

\def\A{{\bf A}}
\def\B{{\bf B}}
\def\x{{\bf x}}

\def\y{{\bf y}}
\def\k{{\bf k}}

\def\D{{\bf D}}

\def\p{{\bf p}}
\def\E{{\bf E}}
\newcommand{\bdel} {{\mbox{\boldmath $\nabla$}}}

\begin{document}
\voffset = 0.3 true in
\topmargin = -1.0 true in 

\title{A Coulomb Gauge Model of Mesons}

\author{Norbert Ligterink\footnote{Current address:University of Twente, Control Engineering
PO Box 217, 7500 AE Enschede, The Netherlands}
}
\affiliation{
Department of Physics and Astronomy, University of Pittsburgh,
Pittsburgh PA 15260}

\author{Eric S. Swanson}
\affiliation{
Department of Physics and Astronomy, University of Pittsburgh,
Pittsburgh PA 15260}
\affiliation{
Jefferson Lab, 12000 Jefferson Ave,
Newport News, VA 23606}

\vskip .5 true cm
\begin{abstract}
A model of mesons which is based on the QCD Hamiltonian in Coulomb
gauge is presented. The model relies on a novel quasiparticle basis to
improve the reliability of the Fock space expansion. It is also 
relativistic, yields chiral pions, and is tightly constrained by QCD 
(quark masses are the only parameters). 
Applications to hidden flavor mesons yield
results which are comparable to phenomenological constituent quark 
models while revealing the limitations of such models.
\end{abstract}

\maketitle

\section{Introduction}

The successes of the quark model of the 1960's led directly to 
the development of QCD in the early 1970's.  A central feature of
the early quark model was the use of constituent quarks as
the relevant degrees of freedom of matter fields. Although the 
advent of QCD changed the details -- the ``light" constituent
quarks of Copley, Karl, and Obryk have become standard and one gluon
exchange is typically employed to describe short 
range dynamics -- the concept of constituent quarks has remained productive 
and pervasive.

QCD also indicates where the quark model may fail. The canonical 
nonrelativistic quark model relies on a potential description of
quark dynamics and therefore neglects many-body effects in QCD. 
Related to this is the question of the reliability of nonrelativistic
approximations, the importance of hadronic decays, and the chiral 
nature of the pion.  The latter two phenomena depend 
on the behavior of nonperturbative glue and as such are crucial to
the development of robust models of QCD and to understanding soft gluodynamics.
Certainly,  one expects that gluodynamics will make its presence felt with
increasing insistence as experiments 
probe higher excitations in the spectrum. Similarly the chiral nature of
the pion cannot be understood in a fixed particle number formalism.  This
additional complexity is the reason so few models attempt to derive the
chiral properties of the pion.  This is an unfortunate situation since
the pion is central to much of hadronic and nuclear physics.

To make progress one must either resort to numerical experiments or 
construct models which are closer to QCD. Here we present one such model
which is based on the QCD Hamiltonian in Coulomb gauge. The Hamiltonian
approach is appropriate for an examination of the bound state problem
because the familiar machinery of quantum  mechanics may be employed and
because all degrees of freedom are physical in Coulomb gauge. Furthermore,
an explicit time-independent potential exists which permits the construction
of bound states in a fixed Fock sector.
The model consists of a truncation of QCD to a set of diagrams which 
capture the infrared dynamics of the theory. The efficiency of the
truncation is enhanced through the use of quasiparticle degrees of
freedom, as will be explained subsequently. Finally, the random phase
approximation (RPA) is used to obtain mesons. This many-body truncation is 
sufficiently powerful to generate Goldstone bosons and has the
advantage of being a relativistic truncation of QCD\cite{RS}.

Because the Hamiltonian is derived from a local density it is covariant,
although the use of Coulomb gauge hides this. The truncations
which we will employ do not ruin this property.
We remark that covariance requires a combination of boost
and gauge transformations in noncovariant gauges
and therefore some care must be taken in the
computation of quantities such as form factors or heavy meson
decay rates. Here we focus on static meson properties 
in the rest frame where these issues do not arise. Finally, we
note that maintaining relativistic invariance in schemes which
extend the RPA may be difficult because the interaction is no longer instantaneous
at higher order. Thus 
different terms must be summed to yield covariant
results and amplitudes may arise which do not have simple wavefunction, 
or RPA, analogues such as amplitudes with a mixture of forward and backward 
moving particles.

Using a single framework to generate chiral
symmetry breaking and the meson spectrum consistently
has been attempted before. LeYaouanc {\it et al.}
\cite{LeYaouanc:1984dr} solved a simple gap equation with a
quadratic interaction and then used the RPA approximation to
obtain chiral pions. Although the interaction is unrealistic
it allowed the important simplification of turning integral
equations into simple differential equations.
Llanes-Estrada and Cotanch \cite{LC} also studied
low-lying states with a linear potential while ignoring
state mixing.
Neither paper considered the effects of the one gluon exchange potential or
renormalization. 

An extensive literature on relativistic quark
models (which do not consider chiral symmetry breaking) exists.
For example, a preliminary study of Fock sector mixing in a
relativistic quark model was performed by Koniuk and Zhang\cite{ZK}.
Detailed examinations of meson and baryon properties have been carried
out by the Bonn group\cite{bonn} in a Salpeter equation framework with
a model confinement potential and Dirac structure. Similarly, extensive
computations of pion and kaon properties have been carried out in 
a covariant Euclidean space Dyson-Schwinger formalism\cite{MR}.

Finally, the gluonic portion of the formalism presented here has been 
used to derive the quenched 
positive charge conjugation glueball spectrum\cite{ss8}. The results are
in very good agreement with lattice computations, indicating that the 
method has some promise.

\section{Model Definition: the Quark Vacuum and Chiral Symmetry Breaking}

Generating the meson spectrum proceeds in three steps: (1) a quasiparticle
basis for the gluonic sector of QCD is obtained with standard many-body 
techniques, (2) this procedure yields an instantaneous interaction which 
is used to construct a quasiparticle (constituent) basis in the quark 
sector, (3) bound state properties are obtained in the the random phase 
approximation.
The first step contains an important
complication: the quasiparticle interaction of QCD depends on the
quasiparticles themselves (in a way made clear below) and hence must
be solved along with the gap equation. This allows the possibility of
deriving the constituent quark interaction if one can obtain the
functional form of the interaction.  We note that this is similar to
solving coupled Dyson-Schwinger equations for, say, the gluon propagator
and vertices.

\subsection{QCD in Coulomb Gauge}

The Coulomb gauge QCD Hamiltonian may be written as $H_{QCD} = H_0 + H_{int}$
with\cite{CL,schwinger, ss7}

\begin{equation}
H_0 = \int \psi^\dagger \left( - i \bm{\alpha} \cdot \bdel + \beta m\right)
\psi + {1 \over 2}\int d\x  \left( \bm{\Pi}^2 - \A \bdel^2 \A \right)
+ {1\over 2}\int d\x d\y\, \rho^a(\x) K^{(0)}(\x-\y) \rho^a(\y) \ \ .
\label{h0}
\end{equation}
The interaction term contains the familiar transverse gluon color charge 
interaction and all of the higher order terms 
due to the non-Abelian nature of QCD:
\begin{eqnarray}
H_{int} &= {1\over 2} \int d\x \left[ \B^2 + \A \bdel^2 \A \right]  -
g \int \psi^\dagger \bm{\sigma}\cdot \A \psi + V_A + V_B +\nonumber \\
& +  {1\over 2}\int d\x d\y\, \rho^a(\x) \left[ K^{ab}(\x-\y;\A) - \delta^{ab}K^{(0)}(\x-\y)
\right] \rho^b(\y)
\label{hint}
\end{eqnarray}
The density $\rho^a$ entering these equations is the full color charge due 
to quarks and gluons

\begin{equation}
\rho^a({\bf x}) =
 f^{abc} {\bf A}^b({\bf x}) \cdot \bm{\Pi}^c({\bf x}) +
 \psi^{\dag}(\x){\lambda^a\over 2}\psi(\x).
\label{rho}
\end{equation}
The instantaneous non-Abelian Coulomb interaction in Eq. \ref{hint} 
is given by 

\begin{equation}
K^{ab}({\bf x},{\bf y};\A) \equiv \langle{\bf x},a|
 { g \over { \bdel\cdot {\bf D}}}(-\bdel^2)
 { g \over { \bdel\cdot{\bf D}}}|{\bf y},b\rangle,
\label{coulk}
\end{equation}
where $\D$ is the covariant derivative in adjoint representation
\begin{equation}
\D^{ab} = \delta^{ab}\bdel - g  f^{abc} \A^c.
\end{equation}
The electric and magnetic fields are defined by

\begin{equation}
\Pi^a \equiv -\E_{tr}^a = \dot\A^a + g (1 - \bdel^{-2} \bdel
\bdel\cdot) f^{abc} A^{0b} \A^c
\end{equation}
and
\begin{equation}
\B^a = \bdel \times \A^a + {1\over 2} g f^{abc} \A^b \times \A^c.
\end{equation}

The interaction in $H_0$ is defined as the vacuum expectation value
of the Coulomb interaction

\begin{equation}
\delta ^{ab} K^{(0)}(\x-\y) = \langle \Psi_0 \vert K^{ab}(\x,\y;\A)\vert \Psi_0\rangle.
\label{k0}
\end{equation}
The vacuum state will be defined shortly.

Finally, the imposition of Coulomb gauge restricts the theory to a curved gauge
manifold with a metric given by $\langle \Phi| \Psi \rangle = \int {\cal D}\A {\cal J}[\A] \Phi^*(\A)\Psi(\A)$.  The factor ${\cal J}$ is the Faddeev-Popov 
determinant given by 

\begin{equation}
{\cal J} = {\rm det}(\bdel \cdot \D).
\end{equation}

The Faddeev-Popov determinant may be removed from the metric by rescaling the
Hamiltonian

$$ 
H \to {\cal J}^{1/2} H {\cal J}^{-1/2}
$$
which is hermitian with respect to $( \Phi|\Psi ) = \int {\cal D\A}
\Phi^*(\A) \Psi(\A)$.  We note that  the nontrivial metric induces 
two new terms, denoted 
$V_A$ and $V_B$, which correspond to the anomalous terms of
Schwinger\cite{schwinger,CL}.

It has been speculated that confinement is related to the well-known
Gribov ambiguity\cite{gribov,Z}. The Gribov ambiguity arises because the existence
of topologically inequivalent solutions to the gauge condition $\bdel \cdot \A = 0$ implies that the gauge is incompletely specified. 
Gribov proposed that the ambiguities with Coulomb gauge may be
resolved by considering fields with $\det(\bdel\cdot\A) \geq 0$ which 
comprise the `Gribov region' (GR). We note that the appearance of the inverse of
the Faddeev-Popov operator in the instantaneous Coulomb potential implies
that gauge configurations near the boundary of the Gribov region create a
strong infrared enhancement in the interaction. It is possible that this
enhancement is the origin of confinement in Coulomb gauge\cite{Z}.

Subsequent research has shown that the Gribov region actually contains gauge
copies and hence gauge fields must lie in a subset of the Gribov region
called the fundamental modular region (FMR).
The FMR contains no redundant field
configurations and may be defined, for example, as the
set of global minima over gauge transformations of the functional 
\begin{equation}
F_\A [g] = {\rm Tr}\int d^3x (\A^g)^2
\end{equation}
where $\A^g = g\A g^\dagger - g\bdel g^\dagger$. It is now known that the
FMR contains the GR except at certain points where the regions
coincide, the FMR is convex, and the GR contains the origin\cite{vanbaal}. 
Furthermore, Zwanziger has recently argued\cite{Z3} that observables have their
support over the intersection of the FMR and the GR, thereby resurrecting
the infrared enhancement argument of the previous paragraph. Finally,
topological aspects of QCD  may be introduced to the formalism via the 
imposition of wavefunctional boundary
conditions on the boundary of the FMR\cite{vanbaal}.

\subsection{Gluon Gap Equation and the Instantaneous Interaction}

It is clear that the interaction of quarks is strongly influenced by
the instantaneous Coulomb interaction of Eq. \ref{coulk}.  It is known,
moreover, that the instantaneous interaction is renormalization group 
invariant\cite{Z}.
This fact permits a physical interpretation of the instantaneous
potential which is a central aspect of our formalism. Furthermore, Zwanziger
has shown that the
Coulomb interaction provides an upper bound to the Wilson loop potential
and has postulated that this bound is, in fact, saturated\cite{Z1}. We
remark that this saturation is crucial to the formalism presented here.
This postulate may be checked with lattice computations, unfortunately the
results are mixed \cite{ZC,G}. If the evidence to the
contrary is confirmed, this approach must be abandoned.

The nonpolynomial functional dependence of the Coulomb interaction on
the vector potential complicates computations in Coulomb gauge.
We proceed by separating the interaction 
into two parts denoted by $K^{(0)}$ (see Eq. \ref{k0}) and $K-K^{(0)}$.
Since $K^{(0)}$ is a vacuum expectation value it contains all diagrams
in which gluons are attached to the operator $\rho K \rho$.  Thus the
remainder  necessarily contains gluons which propagate. Since gluons
are quasiparticles  with a dynamical mass on the order of 1 GeV (the 
hybrid-meson mass gap), matrix elements of $K-K^{(0)}$ are suppressed
in typical hadronic observables, considerably simplifying computations.

In spite of this, it is impossible to obtain a closed form expression for the
the vacuum expectation value of $K$. However, a procedure for obtaining 
Dyson equations which sum the leading infrared diagrams for this matrix element
has been described in Ref.\cite{ss7}.  
An important element in the formalism is the construction of a basis
which permits an efficient Fock space expansion. The method proposed in 
Ref. \cite{ss7} built this basis with the aid of a gluonic vacuum Ansatz (it is worth stressing that
the vacuum need not be highly accurate, it merely provides a starting point for
constructing a more general quasiparticle basis, and any sufficiently general
vacuum Ansatz will suffice). The Ansatz employed is

\begin{equation}
\Psi_0[\A] = \langle \A | \omega \rangle =  \exp\left[-{1\over 2} \int d\k
   \A^a(\k)\,\omega(k)\A^a(-\k) \right].
\label{varvac}
\end{equation}
which depends on an unknown trial function, $\omega(k)$. This is the simplest
correlation one can build into the vacuum and corresponds to the BCS Ansatz
of many-body physics.
Note that the perturbative vacuum is obtained when $\omega = |\k|$. 

The trial
function is obtained by minimizing the vacuum energy density

\begin{equation}
  {{\delta}\over {\delta \omega}} \langle \omega |H |\omega \rangle  = 0.
  \label{vev}
\end{equation}
The vacuum state obtained from this procedure is denoted $|\omega\rangle$.
We refer to
$\omega$ as the gap function since it is also
responsible for lifting the single particle gluon
energy beyond its perturbative value.
Of course evaluating the matrix element in Eq. \ref{vev} requires an explicit
expression for $\langle\omega |\rho K \rho|\omega\rangle$ which is provided 
by the Dyson procedure described above. The result is a set of coupled 
nonlinear integral equations
which describe the gap equation and the Dyson equation for $K^{(0)}$.  Solving
these equations yields both the quasigluon dispersion relation, $\omega(k)$,
and the quasigluon effective interaction, $K^{(0)}(\x-\y)$. Renormalization
is achieved by fitting the latter
to the lattice Wilson loop potential. The result is in excellent agreement with
the lattice and provides a dynamical mass scale for the quasigluons: $m_g \equiv 
\omega(0) \approx 600$ MeV.  
It is this large mass scale which permits rapid 
convergence of any Fock series expansion in the gluonic sector of QCD and explains
why quark degrees of freedom dominate low energy hadronic physics.
We remark that the emergence of a confining potential
is nontrivial and indicates the robustness of the method.

An analytic approximation to the solution to the coupled equations yields the
following form  for the vacuum expectation value of the Coulomb interaction\cite{ss7}:

\begin{equation}
K^{(0)}(k) = {12.25\over k^2} \left\{ \begin{array}{ll}
                               \left({m_g \over k}\right)^{1.93} & k < m_g\\
                               { 0.6588 \log(k^2/m_g^2 + 0.82)^{-0.62} \log(k^2/m_g^2 + 1.41)^{-0.80}} & k > m_g 
                                      \end{array} \right.
\label{Vss7}
\end{equation}
This form mimics the numerical solution and the lattice results quite well. The
long range potential behaves as $k^{-3.93}$, within 2\% of the expected linear
behavior (the deviation is likely to be a numerical artifact). This allows the
extraction of a string tension via $6 \pi b \approx 12.25 m_g^2$, which implies
$b \approx 0.234$ GeV$^2$. The effect of the smaller exponent is to reduce this
string tension to approximately 0.21 GeV$^2$  at physically relevant scales. A
fit to the standard `Coulomb plus linear' potential (const - $4/3\cdot \alpha_s/r + br$)
yields an effective strong coupling of $\alpha_s = 0.12$. This value is rather
small with respect to expectations (in part because there was little lattice data
at small distances in the fit of Ref.\cite{ss7}). Thus we have also employed a
modified potential of the form

\begin{equation}
K^{(0)}(k) = {12.25\over k^2} \left\{ \begin{array}{ll}
                               \left({ m_g \over k}\right)^{1.93} & k < m_g\\
                               \left({\log(1+a)\over \log(k^2/m_g^2 + a)}\right)^b & k > m_g 
                                      \end{array} \right.
\label{Vmod}
\end{equation}
Here $a$ and $b$ are constants which may be varied. Values of $a = 1.0$ and $b=0.8$ 
give a potential which agrees well with lattice data at short distance, 
but which otherwise is very close to the original form of Eq.\ref{Vss7}.
Both potentials have been employed in the following to test the sensitivity of the 
results on the functional form of the potential.


\subsection{Quark Gap Equation}

Our chief goal is to examine a model of QCD which permits the simultaneous
description of heavy quarkonia and chiral pions.  It is clear that this
may not be achieved in a potential formalism -- the many-body aspect of
QCD is required. Our approach to the quark sector therefore mimics that
of the gluon sector, namely we model the quark vacuum with a Gaussian 
wavefunctional and determine the quark gap equation. Solving this yields
dynamical chiral symmetry breaking and a massless Goldstone boson in the
random phase approximation in accord with the Thouless sum rule\cite{thouless}. 
We note that this general approach has been used many times in the past, starting with 
the classic work of Nambu and Jona-Lasinio \cite{NJL}.  Subsequent work has dealt
with renormalization issues \cite{A,FM} or with models which are closer to 
QCD \cite{Orsay}. How the constituent quark model may be reconciled with chiral
symmetry breaking is explained in Ref. \cite{ss5}. Finally an extensive literature on the Dyson-Schwinger approach to
this problem exists\cite{MR}.

In this approach the gap equation represents a nonperturbative one loop
computation and thus must be properly renormalized. As noted in the gluonic
sector, this is a nontrivial step whose implementation depends on the subset
of diagrams being summed. In the BCS/RPA formalism employed here, we have 
found that standard
renormalization is sufficient to guarantee finite results. In particular we
have added mass renormalization 

$$
 \delta H_m = \delta m\int d\x\, \psi^\dagger \beta \psi 
$$
and wavefunction renormalization
$$
\delta H_\psi = \left( Z_\psi - 1\right) \int d\x\,\psi^\dagger (-i {\bm \alpha}\cdot \bdel) \psi
$$
terms to the Hamiltonian of Eq. \ref{h0} and the theory has been truncated at 
the scale $\Lambda$. Recall that the effective instantaneous interaction
has already been rendered finite.

Proceeding with the standard Bogoliubov or Dyson procedure yields the following
quark gap equation

\begin{equation}
Z_\psi(\Lambda) p s_p = \left[m(\Lambda) + \delta m(\Lambda)\right]c_p + 
{C_F \over 2} \int {k^2 dk \over (2\pi)^3} \left[ V_0(p,k) s_k c_p - V_1(p,k) c_k s_p\right]
\label{qgap}
\end{equation}


\noindent
where 
$$
V_L(p,k) = 2\pi \int d(\hat p \cdot \hat k)\, K^{(0)}(\p-\k) P_L(\hat p \cdot\hat k).
$$
The functions $s_k$ and $c_k$ are defined in terms of the Bogoliubov angle $\phi(k)$
as $s_k = \sin\phi(k)$ and $c-k = \cos \phi(k)$. 
The quark gap equation is to be solved for the unknown Bogoliubov angle, which 
then specifies the quark
vacuum and the quark field mode expansion via spinors of the form

\begin{equation}
u_s(k) = \sqrt{1 + s_k \over 2} \left( \begin{array}{c}
                                     \chi_s  \\
                                      {c_k\over 1 + s_k}  {\bm{\sigma} \cdot \hat k} \chi_s
                                 \end{array} 
                                  \right).
\end{equation}
Comparing the quark spinor
to the canonical spinor (we use nonrelativistic normalization) permits a simple
interpretation of the Bogoliubov angle through the relationship $\mu(k) = k \tan\varphi(k)$ where $\mu$ may 
be interpreted as a dynamical momentum-dependent quark mass. Similarly $\mu(0)$
may be interpreted as a constituent quark mass.

In the case of massless quarks the right hand side of the quark gap equation 
diverges logarithmically for potentials
obeying the perturbative relation $K^{(0)}(k) \to k^{-2}$ for large $k$. The 
divergence 
 may be absorbed
into the wavefunction renormalization, $Z_\psi = 1 - C_F/(6\pi^2) \log\Lambda$,
yielding a finite gap equation.  It is also possible to renormalize by examining
the once-subtracted gap equation. 
For the massive quark case two logarithmic divergences proportional
to the quark mass and momentum appear. It is convenient to absorb these divergences
separately into the mass and wavefunction terms respectively.
For the study presented here the potential is modified by logarithmic corrections 
at short
distances, thus all integrals are finite and the cutoff may be removed immediately.
We note, however, that it still may be useful to make finite renormalizations.

The numerical solution for the dynamical quark mass is very accurately represented
by the functional form

\begin{equation}
\mu(k) = \sigma K^{(0)}(k)\left( 1 - {\rm e}^{-{M /(\sigma K^{(0)}(k)})}\right)
\label{muEq}
\end{equation}
where $M$ is a `constituent' quark mass and $\sigma$ is a parameter related to the
quark condensate. Notice that this form approaches  the constituent mass for small 
momenta and $\sigma K^{(0)}$ for large momenta. The latter behavior is in accord with
the quark gap equation which implies that\cite{P}

\begin{equation}
\mu(k) \to {C_F\over 2} \int {d^3q \over (2\pi)^3} K^{(0)}(k) s(q)
= -{C_F \over 4 N_c} K^{(0)}(k) \langle \bar \psi \psi \rangle.
\label{muk}
\end{equation}

A rough fit to the numerical solution yields $M = 68$ MeV and $\sigma =$ 0.001 GeV$^3$. The constituent quark mass is small compared to typical relativistic
constituent quark model masses of roughly 200 MeV.   The value for $\sigma$ implies
a quark condensate of approximately (-210 MeV)$^3$, in reasonable agreement with
current estimates of (-250 MeV)$^3$. However, we note that direct computations of the condensate 
typically yield results of approximately (-110 MeV)$^3$. 
These flaws undoubtedly point to inadequacies
in the quark vacuum Ansatz. Of course, since we are working with the full QCD
Hamiltonian, it is possible to improve the Ansatz (for a coupled cluster approach,
see Ref. \cite{SK}; for one loop corrections see Ref. \cite{ss4}). Since one of our chief interests is the implementation of a
formalism which respects chiral symmetry, and not detailed numerics, we satisfy
ourselves with the present procedure.


\section{Mesons}

With explicit expressions for the quasigluon interaction (and hence the constituent
quark interaction) and the dynamical quark mass in hand we are ready to obtain mesonic
 bound
states. As mentioned above, in order to obtain a massless pion one must construct states, $|M\rangle$, in the random phase approximation:

\begin{equation}
\langle M| [H,Q^{\dag}_M] | RPA \rangle = (E_M - E_{BCS} ) \langle M| Q^{\dag} |RPA \rangle, 
\label{RPA}
\end{equation} 
where $Q^{\dag}_M$ is defined in terms of the positive and
negative energy wavefunctions, $Q^{\dag}_M = \sum_{\alpha\beta}
\left[ \psi^+_{\alpha\beta} B^{\dag}_{\alpha} D^{\dag}_{\beta} - \psi^-_{\alpha\beta}
D_{\beta} B_{\alpha} \right]$ with $B$ and $D$ being the 
quasiparticle operators. It is worthwhile recalling that the RPA  method is
equivalent to the Bethe-Salpeter approach with instantaneous interactions\cite{RS}.

The RPA and TDA equations includes self energy terms (denoted $\Sigma$)  for each quark line and these must be
renormalized.
In the zero quark mass case renormalization of the TDA or RPA equations
proceeds in the same way as for the quark gap equation. In fact, the
renormalization of these equations is consistent and one may show that a 
finite gap equation implies a finite RDA or TDA equation. This feature
remains true in the massive case. 

The RPA equation in the pion channel reads:

\begin{eqnarray}
(E_\pi-E_{BCS}) \psi^+(k) &=& 2\left[m s_k + k c_k + \Sigma(k)\right] \psi^+(k) \nonumber \\
&& - {C_F\over 2} \int {p^2dp\over (2\pi)^3}\left[ V_0(k,p)(1+s_ks_p) + V_1(k,p)c_kc_p\right] \psi^+(p) \nonumber \\
&& - {C_F\over 2} \int {p^2dp\over (2\pi)^3}\left[ V_0(k,p)(1-s_ks_p) - V_1(k,p)c_kc_p\right] \psi^-(p)
\end{eqnarray}
A similar equation for $\psi^-$ holds with $(+ \leftrightarrow -)$ and $E \to -E$. The
wavefunctions $\psi^\pm$ represent forward and backward moving components of the
many-body wavefunction and the pion itself is a collective excitation with infinitely
many constituent quarks in the Fock space expansion.

We also consider a simpler truncation of QCD called the Tamm-Dancoff approximation.
This may be obtained from the RPA equation by neglecting the backward wavefunction
$\psi^-$.

We have computed the spectrum in the random phase and Tamm-Dancoff approximations
and confirm that the pion is massless in the chiral limit.  We also find that 
the Tamm-Dancoff approximation yields results very close to the RPA for all states
except the pion: the TDA pion mass is 580 MeV while the first excited pion 
has an RPA mass of 1410 MeV  and a TDA mass of 1450 MeV. All other mesons have nearly
identical RPA and TDA masses.
For this reason we simply present  TDA equations and results below.

The complete hidden flavor meson spectrum in the Tamm-Dancoff approximation is
given by the following equations.

\begin{equation}
E \psi_{PC}(k) = 2 \left[m s_k + kc_k + \Sigma(k)\right]\psi_{PC}(k) - {C_F\over 2} \int {p^2dp\over (2\pi)^3} K^{PC}_J(k,p) \psi_{PC}(p)
\end{equation}
with

\begin{equation}
\Sigma(k) = {C_F\over 2} \int {p^2 dp\over (2\pi)^3} \left( V_0 s_k  s_p + V_1 c_k c_p \right)
\end{equation}
and where $\psi$ is the meson radial wavefunction in momentum space.

An alternate form for the kinetic and self energies which is closer the Schr\"odinger
equation may be obtained by substituting the gap equation to obtain:

\begin{equation}
\mbox{ kinetic + self energy} = 2\left[ E(k) + \Gamma(k)\right]
\end{equation}
where
\begin{equation}
\Gamma(k) = {C_F \over 2} \int {p^2 dp \over (2\pi)^3} V_1 {c_p\over c_k}.
\end{equation}
and
\begin{equation}
E(k) = \sqrt{k^2+\mu(k)^2}.
\end{equation}

The kernel, $K_J$ in the potential term depends on the meson quantum numbers, $J^{PC}$.
In the following possible values for the parity or charge conjugation eigenvalues are
denoted by $(J) = +$ if $J$ is even and $-$ if $J$ is odd.

\noindent
$\bullet\ $ $0^{++}$

\begin{equation}
K(p,k) = V_0 c_p c_k + V_1 (1 + s_p s_k) 
\end{equation}

\noindent
$\bullet\ $ $J^{(J+1)(J)}$ [$^1J_J$ , $J \geq 0$]

\begin{equation}
K_J(p,k) = V_J (1 + s_p s_k) + \left( V_{J-1} {J\over 2J+1} + 
V_{J+1} {J+1 \over 2J+1} \right) c_p c_k
\label{1jj}
\end{equation}

\noindent
$\bullet\ $ $J^{(J+1)(J+1)}$ [$^3J_J$ , $J \geq 1$]

\begin{equation}
K_J(p,k) = V_J (1 +s_p s_k) + \left( V_{J-1} {J+1\over 2J+1} 
+ V_{J+1} {J \over 2J +1} \right) c_p c_k
\end{equation}

\noindent
$\bullet\ $ $J^{(J)(J)}$ [$^3(J-1)_J, ^3(J+1)_J$ , $J \geq 1$]

\begin{eqnarray}
K_{11}(p,k) &=& V_J c_p c_k + \left( V_{J-1} {J \over 2J+1} + V_{J+1} 
{J+1\over 2J+1} \right) (1 + s_p s_k) \cr
K_{22}(p,k) &=& V_J c_p c_k + \left( V_{J-1} {J+1 \over 2J+1} + V_{J+1} 
{J\over 2J+1} \right) (1 + s_p s_k) \cr
K_{12}(p,k) &=& \left( V_{J-1} - V_{J+1} \right)
{\sqrt{J(J+1)}\over 2J+1} \left( s_k + s_p \right)
\end{eqnarray}

These interaction kernels have been derived in the quark helicity basis.
We remark that in the LS basis the off-diagonal (J-1):(J+1) interaction 
for $J^{(J)(J)}$ mesons 
is proportional to $1 + s_k s_p - s_k - s_p$ and hence goes to zero in
the heavy quark mass limit as expected. Finally, we note that the authors
of Ref. \cite{LC} find that the $1^{++}$ and $1^{+-}$ kernels are identical.
The likely reason is an error in their $1^{+-}$ kernel which disagrees with 
that of Eq. \ref{1jj}.

\subsection{Short Range Behavior}

Our quark interaction is instantaneous, central, and obeys Casimir scaling.
However, it is not flavor or spin independent because the full spinor structure of 
the interaction has been retained. This spinor structure is specified by the
Hamiltonian of QCD and is of a vector nature (specifically $\bar \psi \gamma_0 \psi
\otimes \bar \psi \gamma_0 \psi$). 
A nonrelativistic reduction of this interaction
yields no spin-spin hyperfine or tensor interactions. Thus the present computation
cannot correctly
describe well known spin splittings such as $J/\psi - \eta_c$ or $\Delta-N$
(we describe the extensions necessary to do so below).

A spin-orbit interaction is present and is given by 

\begin{equation}
V_{SO} = {1 \over 2 m^2 r} {d \over dr}\left( -C_F K^{(0)}(r)\right) \vec L \cdot \vec S
\end{equation}
in the equal quark mass case.

The spin-orbit interaction is famous for its problematic nature in the
constituent quark model. Quark model lore states that the spin-orbit interaction
generated by one gluon exchange is too strong for phenomenology and must be 
softened by the addition of a spin orbit interaction from the confinement
term which has an opposite sign. This can only be arranged if confinement
has a scalar Dirac structure ($\bar \psi \psi \otimes \bar \psi \psi$). 
This is clearly at odds with the formalism presented here which insists that
the confinement and Coulomb potentials (here we mean the perturbative $1/r$ tail of the 
static potential)  share the same Dirac structure. The
resolution to the conundrum is  that it is too simple-minded to ascribe all
of spin-orbit interactions to the Dirac structure of instantaneous potentials.
Indeed, spin-dependence in QCD is partly generated by nonperturbative
mixing with intermediate states, and need not follow the dictates of quark
model lore. A specific  realization of this is given in Ref. \cite{ss3}.
Finally, we note that a scalar confinement interaction
leads to inconsistencies between mesons and baryons: if mesons confine then
baryons anticonfine --- clearly an unacceptable situation!

Relativistic interactions generate short range spin-dependent interactions
by virtue of their spinor structure. The other sources of spin-dependence are
topological effects (for example, an instanton induced interaction) and Fock
sector mixing effects. Fock sector mixing is easily incorporated into the
current formalism, one need only increase the size of the Fock space being
considered (there is one subtlety: the mixing between Fock sectors must be
treated carefully since it can involve nonperturbative gluodynamics). One expects
the leading higher Fock sectors to be meson-meson (ie., meson loop corrections to 
the spectrum) and hybrid. The latter case is the nonperturbative analogue of
one-gluon-exchange, and as mentioned above, is phenomenologically important
in the light meson spectrum. 

We shall leave the topic of Fock space mixing for a future investigation and
press ahead with an examination of the spectrum which arises from the central
static potential generated by the non-Abelian Coulomb interaction, keeping in mind
that large spin-dependent mass shifts may occur in the light spectrum.

\section{Results and Discussion}

Once the gluonic sector of the formalism has been fixed by renormalization 
the only remaining parameters are quark masses. In the following we have determined
these by fitting the $\Upsilon$, $J/\psi$, and $\phi$ masses. We work in
the chiral limit for light ($u$ and $d$ quark) mesons so predictions in this
sector are completely fixed.  As a result there is no possibility of adjusting
the spectrum presented below. Thus it is possible to test the assumptions of
the model throughout the spectrum.

\subsection{Quarkonia}

A simple way to obtain the upsilon spectrum from QCD is to insert the
lattice Wilson potential into the nonrelativistic Schr\"odinger equation.
One argues that the heavy $b$ quark mass validates the Born-Oppenheimer
approximation (so that the static potential may be used) and the use
of a  nonrelativistic framework. Precisely this approach has been taken by 
Juge, Kuti, and Morningstar\cite{JKM} (JKM)
and was subsequently justified by comparison 
with nonrelativistic lattice computations\cite{JKM2}.

Since the 
potential we employ is essentially equivalent to that obtained from the
Wilson loop 
one may expect that the predicted upsilon spectrum will agree very closely 
with that of JKM. This expectation relies on two 
things: (i) the heavy quark mass must eliminate non-central contributions 
in the interaction kernel, (ii) the self energy contribution must be essentially
independent of momentum. (The authors of Ref. \cite{JKM} did not
include a self energy term in their model because it cannot be extracted from lattice
data. Such a term must exist as external fermion propagator  corrections.)  
Explicit computations show that these expectations are indeed borne out and indicate
that a finite renormalization is sufficient to largely eliminate the self energy term.

Our predicted upsilon spectrum is given in Table I.  
A comparison with experiment
shows that the radial excitations rise too slowly with radial quantum number, and
indeed our results similarly disagree with those of JKM. But, as noted above, the
form for the static potential of Eq. \ref{Vss7} is not strongly constrained at
large momenta. We have therefore computed the upsilon spectrum with the modified
potential of Eq. \ref{Vmod}. The results are shown in Table II and are in much
better agreement with JKM and experiment. As expected, the detailed form of the
Coulomb tail of the potential is very important for low-lying heavy quark states. 
Our results are very similar to typical quark model spectra\cite{GI}; with some deviation
(at the percent level) becoming visible higher in the spectrum. This is to 
be expected because we employ a lattice potential with a string tension of 0.2-0.25 
GeV$^2$ whereas the quark model of Ref. \cite{GI} takes $b = 0.18$ GeV$^2$. 

Overall, the agreement with the experimental upsilon spectrum is impressive considering
the simplicity of the model and that the potential was not fit to data. It is possible
that deviations are seen higher in the spectrum, and indeed, one may expect this
since the open flavor threshold is at 10.56 GeV -- between the third and fourth
vector S-wave states. In general mixing with higher Fock components, such as hybrids (one 
gluon exchange in perturbative language) and meson-meson channels, will occur. These
effects should become more important as one probes higher in the spectrum. 
Eventually the quark + potential picture of mesons should break down entirely as
an increasing number of gluonic degrees of freedom are excited.

The psi spectrum predicted with the canonical potential of Eq. \ref{Vss7} is presented
in Table III. The agreement with experiment is on the percent level. This is something
of a surprise when considering that the upsilon spectrum required a careful fit to
the high momentum potential. The psi spectrum computed from the modified
potential (Table IV) compares unfavorably to the data, with deviations at
the five percent level. It appears that the linear and Coulomb portions of the
potential are equally important to the low-lying psi spectrum and that the small
value of the effective strong coupling has largely cancelled against the large
value of the string tension. Indeed, as mentioned above, quark models typically employ a much smaller
string tension than that of the Wilson loop. We thus have evidence that the
effective string tension is reduced for lighter quark masses. This can easily
be induced by mixing with virtual meson pairs or by motion of the sources in the
Wilson loop. Again, it should be possible to incorporate the physics of string
softening in the model through Fock sector mixing.

The results of Tables III and IV also indicate that spin splittings are becoming
important. For example the $\eta_c$ is roughly 100 MeV lighter than the $J/\psi$
and this is not predicted in the model. Again, this fault is easy to remedy once
higher Fock components are admitted.  One also sees evidence for tensor splitting in
the $J^{++}$ states which are quite well reproduced with the modified potential (Table IV). One concludes that the Dirac structure of the interaction is becoming important
and that it does a reasonable job at charm mass scales but that Fock mixing must be
accounted for, even for low-lying states.  As indicated in the Introduction, if
canonical quark model lore holds, the addition of virtual hybrid states to the
model will ruin the predicted tensor splittings since these will increase them
an unacceptable amount. We stress, however, that nonperturbative mixing with virtual
hybrid states is not equivalent to perturbative gluon exchange which 
only becomes relevant in Hamiltonian QCD very high in the spectrum.

\subsection{Isovector  Mesons}

We have seen that the formalism presented here is quite accurate for
bottom quarks and reasonably accurate (with the possibility of being very accurate
once simple Fock sector mixing is included) for charm quarks. 
The challenge in carrying this success to the light quark sector lies in 
(i) the assumption that the static Wilson loop potential is relevant to light
quarks, (ii) the importance of chiral symmetry breaking, (iii) large spin-dependences which may be present, (iv) the possibility that topological aspects of QCD become
important. One of the reasons Coulomb gauge QCD is useful for constructing models
of hadronic physics is that these issues may be addressed in a systematic fashion.
The latter three issues have already been discussed; for the first we note that
an instantaneous interaction between quarks exists for all quark masses in Coulomb gauge.
The viability of this interaction can only be affected by higher order gluonic terms
(such as in the operator $K-K^{(0)}$); but we have seen that these contributions  are 
suppressed by an energy denominator of the order of 1 GeV. Thus a static interaction
should provide a good approximation low in the light spectrum. Virtual light quark
loops are another source of nonpotential interactions. But chiral symmetry
breaking implies that the light current quarks acquire an effective mass, and this
mass assists in dampening such loop effects. The success of the constituent quark
model also indicates that loop effects
on the interaction can be largely subsumed by renormalization.

Our results for the light meson spectrum are presented in Figure 1 and in Table V.
The first feature to notice is that the pion is massless as desired. Of course
a finite pion mass can be obtained if a finite current quark mass is used in the
calculation. We have chosen to keep the current quark mass at zero in order to 
test the robustness of the model in a zero parameter computation.

The rho meson mass is predicted to be 772 MeV, in very good agreement with data. 
We regard this as somewhat fortuitous since the potential has been fixed by lattice
data and no parameter tuning has taken place. The first radial excitation is at 1390
MeV, whereas the experimental mass is approximately 1450 MeV. However, we note that
 already the possibility
of Fock mixing arises since the lowest mass vector hybrid is expected around 1900 MeV.

As seen in the psi spectrum, the tensor $J^{++}$ multiplet is sensitive to short
range effects. In this case the $1^{++}$ state is in rough agreement with experiment.
However the tensor state lies much above the data. The isoscalar scalar
has a mass of 850 MeV. One may be tempted to ascribe this state to the
$f_0(980)$, which is famous for being too light in constituent quark models. However,
the fact that the tensor state is much too massive is an indication that additional
strong spin-orbit forces are required to obtain a satisfactory description of light
mesons and that any conclusions concerning the nature of the scalar meson would
be premature.

The figure and table also present meson masses for high angular momentum. It is
seen that the model severely overestimates  the masses of these states. The likely
cause for this is the large (compared to quark model) string tension which we
have used.  However, as argued above, there is little freedom in choosing the
string tension and one must ascribe the discrepancy to nonpotential effects.
The simplest such effects are extra quasigluons in the Fock space expansion of
higher lying states. Thus we regard the poor quality of the predictions at high
angular momentum as an indication of the range of validity of the potential
approach to hadronic physics (of course it is possible to extend this range
by allowing more parameter freedom, but the degrees of freedom will be incorrect,
and detailed predictions of such models must fail).

Equations (27) and (28)  make it clear that meson masses form charge conjugation
doublets in the pattern $1^{+\pm}$, $2^{-\pm}$, {\it etc.} in the large angular momentum
limit.  This pattern is clearly seen in the figure. Such behavior is expected in the
heavy quark limit where quark spins decouple and one may expect it to also occur
for light quarks in the high angular momentum regime since the quark spin effects
are local in the interaction.

A central feature of chiral symmetry breaking is that quark spinors reduce to massless
spinors at momenta much larger than the chiral symmetry breaking scale. Thus one
has $c_k \to 1$ and $s_k \to 0$ for $k >> \Lambda_{\chi SB}$. Since the average 
quark momentum becomes large as the angular momentum increases we conclude that the
interaction kernel approaches 

\begin{equation}
K_J \to V_J + {1\over 2}V_{J-1} + {1\over 2} V_{J+1}
\end{equation}
for all possible $J^{PC}$ (the off-diagonal potential in the $J^{(J)(J)}$ sector
goes to zero). Thus the entire spectrum becomes degenerate in the large angular momentum
limit; in particular parity doubling occurs, as expected\cite{CG}. 

\begin{figure}
\includegraphics[width=14cm]{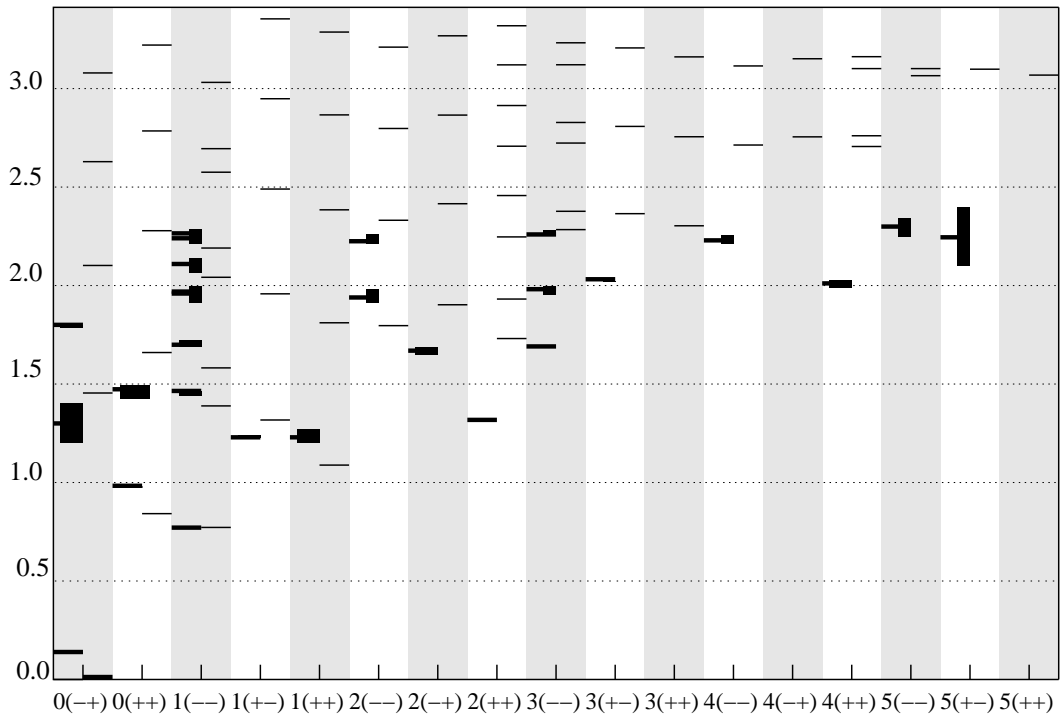}
\caption{ 
\label{uu} 
The isovector meson spectrum.  Experimental masses\cite{PDG} 
are represented by lines to the left of each column. The attached boxes indicate
the widths of each state. Lines to the right are the results of this computation
in the massless quark limit. Narrower boxes are new results from the Crystal Barrel
collaboration\cite{CB}.
}
\end{figure}

\section{Conclusions}

It is becoming clear that simple constituent quark models are limited in their
ability to describe excitations high in the hadronic spectrum. Well known 
flaws like nonrelativistic kinematics are exacerbated by the limitations of
a fixed particle number formulation. Indeed one expects Fock sector mixing to become
increasingly important as experiment probes high in the spectrum. For example 
meson loop effects can cause significant mass shifts and commensurate 
wavefunction distortion will affect other matrix elements. Furthermore, gluons 
must manifest themselves about 1 GeV above the ground state in a given sector and
the subsequent state mixing can be important.

The model constructed here is an attempt at going beyond the nonrelativistic quark model.
Since it is based on a truncation of QCD in Coulomb gauge, it is heavily constrained, 
systematically improvable, and relativistic. Because the instantaneous interaction is 
fixed by the lattice gauge Wilson loop potential the model can also fail.
Any such failure may be regarded as an indication of the limitations of the potential
approach to hadronic physics. Furthermore, the Dirac structure of the interaction is
fixed by the Coulomb gauge Hamiltonian of QCD.  This has important phenomenological 
implications; for example the spin-orbit interaction is fixed to have the same sign
as the Coulomb interaction of the constituent quark model. A happy consequence is that
mesons and baryons are treated on equal footing (scalar confinement requires an 
additional sign change of the interaction between the sectors).  Finally, the assumption
that the operator $K$ generates the leading quark and gluon interaction can be tested
on the lattice. Initial results are mixed, with Ref. \cite{ZC} supporting the conjecture
while Ref. \cite{G}  finds that the Coulomb string tension is roughly three times larger
than the Wilson loop string tension. We find the latter result improbable since the 
coupling of gluons with static quarks is suppressed, leaving only the non-Abelian
Coulomb interaction to mediate the Wilson loop interaction. Nevertheless, if the 
latter result is confirmed the present method will likely have to be abandoned.

The issue of Fock sector mixing will be vital to the success of this program. Such
effects are clearly needed in the light quark spectrum and are of some significance
for heavy quarks. The most important such effect is the nonperturbative analogue of 
one gluon exchange, namely mixing with virtual hybrid states. Implementing this will
be a crucial test of the model since spin orbit splittings depend sensitively in the
Dirac structure of the mixing terms and of the central potential. It is entirely possible
that the formalism proposed here will fail and that some sort of string approach will
be required but this remains to be seen.

We have implemented chiral symmetry breaking using
standard many-body or Nambu--Jona-Lasinio methods. This is of course a truncation of
all diagrams which contribute to the bound state Bethe-Salpeter equation, however, it
is convenient and is enough to  demonstrate the chiral nature of the pion and the
importance of chiral symmetry high in the spectrum. It is possible to improve the
computation in a systematic fashion. Analogous improvements in the Dyson-Schwinger
Bethe-Salpeter approach are discussed in Ref. \cite{RR}.

In future we intend to examine the open flavor spectrum, strong decays, short range
structure and spin splittings, and Fock sector mixing effects.  Further research into
topological aspects of the model and the isoscalar sector will also be of great interest.

\begin{acknowledgments}

This work was supported by the DOE under contracts DE-FG02-00ER41135  and  DE-AC05-84ER40150.
\end{acknowledgments}

\appendix

\begin{table}
\caption{$\Upsilon$ Spectrum (MeV)}
\begin{tabular}{rrrrrrrrrrrrrrrrr}
\hline
 $0^{-+}$& $0^{++}$& $1^{--}$& $1^{+-}$& $1^{++}$&
 $2^{--}$& $2^{-+}$& $2^{++}$& $3^{--}$& $3^{+-}$&
 $3^{++}$& $4^{--}$& $4^{-+}$& $4^{++}$& $5^{--}$&
 $5^{+-}$& $5^{++}$\\
\hline
 9460 &  9723 &  9460 &  9731 &  9727 &  9946 &  9948 &  9735 &  9954 &
10141
  &  10139 &  10318 &  10319 &  10147 &  10326 &  10487 &  10486\\
 9878 &  10070 &  9878 &  10076 &  10073 &  10254 &  10256 &  10079 &
10261 &
 10426 &  10424 &  10586 &  10587 &  10431 &  10593 &  10744 &  10743\\
 10205 &  10369 &  9941 &  10375 &  10372 &  10536 &  10538 &  10133 &
10311
  &  10696 &  10694 &  10850 &  10851 &  10478 &  10639 &  11003 &
11002\\
 10494 &  10646 &  10205 &  10651 &  10649 &  10804 &  10806 &  10378 &
10541
  &  10957 &  10955 &  11103 &  11104 &  10701 &  10856 &  11247 &
11246\\
 10761 &  10901 &  10250 &  10908 &  10904 &  11050 &  11052 &  10419 &
10580
  &  11194 &  11192 &  11336 &  11337 &  10736 &  10889 &  11483 &
11482 \\
\hline
\end{tabular}
\end{table}

\begin{table}
\caption{$\Upsilon$ Spectrum. Modified Potential, $a=1.0$, $b=0.8$ Experiment in brackets. (GeV)}
\begin{tabular}{lllll}
\hline
$0^{-+}$ & $0^{++}$ & $1^{--}$\footnote{S-wave only.} & $1^{++}$ & $2^{++}$ \\
\hline
9.46  & 9.80 (9.86) & 9.46 (fit)     & 9.81 (9.89)   & 9.83 (9.91) \\
9.97 &  10.20 (10.23) & 9.98 (10.02) & 10.21 (10.25) & 10.22 (10.27) \\
10.35 &  10.54        & 10.35 (10.35) & 10.54        & 10.29 \\
10.67 & 10.83         & 10.67 (10.58)   & 10.84      & 10.55 \\
10.96 & 11.10         & 10.95 (10.86?) & 11.10       & 10.60 \\
11.22 & 11.35         & 11.21 (11.02?) & 11.35       & 10.85 \\
\hline
\end{tabular}
\end{table}

\begin{table}
\caption{$\psi$ Spectrum (MeV)}
\begin{tabular}{rrrrrrrrrrrrrrrrr}
\hline
 $0^{-+}$& $0^{++}$& $1^{--}$& $1^{+-}$& $1^{++}$&
 $2^{--}$& $2^{-+}$& $2^{++}$& $3^{--}$& $3^{+-}$&
 $3^{++}$& $4^{--}$& $4^{-+}$& $4^{++}$& $5^{--}$&
 $5^{+-}$& $5^{++}$\\
\hline
 3061 &  3376 &  3063 &  3424 &  3400 &  3720 &  3736 &  3450 &  3772 &
4016
  &  4003 &  4264 &  4274 &  4058 &  4321 &  4517 &  4508\\
 3639 &  3878 &  3642 &  3911 &  3895 &  4154 &  4166 &  3932 &  4196 &
4406
  &  4396 &  4625 &  4633 &  4441 &  4673 &  4851 &  4843\\
 4093 &  4294 &  3684 &  4320 &  4307 &  4531 &  4541 &  3962 &  4218 &
4753
  &  4745 &  4951 &  4958 &  4458 &  4687 &  5157 &  5150\\
 4480 &  4658 &  4096 &  4679 &  4668 &  4868 &  4876 &  4337 &  4566 &
5068
  &  5061 &  5250 &  5256 &  4784 &  4993 &  5440 &  5434\\
 4823 &  4985 &  4126 &  5003 &  4994 &  5175 &  5183 &  4362 &  4586 &
5360
  &  5354 &  5530 &  5536 &  4800 &  5007 &  5710 &  5705\\
\hline
\end{tabular}
\end{table}

\begin{table}
\caption{$\psi$ Spectrum. Modified Potential, $a=1.0$, $b=0.8$ Experiment in brackets. (GeV)}
\begin{tabular}{lllll}
\hline
$0^{-+}$ & $0^{++}$ & $1^{--}$ & $1^{++}$ & $2^{++}$ \\
\hline
3.09 (2.98)  & 3.45 (3.41)&  3.095 (fit) & 3.48 (3.51) & 3.57 (3.55) \\
3.76 (3.65) & 4.03        & 3.76 (3.686) & 4.05        & 4.11 \\
4.28        & 4.51       & 3.81 (3.77) & 4.53          & 4.12\\
4.73        & 4.93       & 4.29 (4.04) & 4.94          & 4.57 \\
5.12        & 5.30       & 4.32 (4.16) & 5.31          & 4.58 \\
\hline
\end{tabular}
\end{table}

\begin{table}
\caption{Isovector Spectrum (MeV)}
\begin{tabular}{rrrrrrrrrrrrrrrrrrrr}
\hline
 $0^{-+}$& $0^{++}$& $1^{--}$& $1^{+-}$& $1^{++}$&
 $2^{--}$& $2^{-+}$& $2^{++}$& $3^{--}$& $3^{+-}$&
 $3^{++}$& $4^{--}$& $4^{-+}$& $4^{++}$& $5^{--}$&
 $5^{+-}$& $5^{++}$& $6^{--}$& $6^{-+}$& $6^{++}$ \\
\hline
 0 &  842 &  772 &  1317 &  1088 &  1796 &  1902 &  1731 &  2284 &
2365 &
 2303 &  2713 &  2755 &  2706 &  3065 &  3098 &  3068 &  3387 &  3410 &
3386\\
 1454 &  1660 &  1389 &  1958 &  1811 &  2331 &  2415 &  1931 &  2377 &
2808
  &  2755 &  3115 &  3151 &  2760 &  3101 &  3461 &  3434 &  3725 &
3746 &
 3412\\
 2102 &  2278 &  1582 &  2490 &  2384 &  2797 &  2865 &  2247 &  2723 &
3206
  &  3161 &  3481 &  3514 &  3101 &  3427 &  3796 &  3771 &  4040 &
4059 &
 3721\\
 2629 &  2785 &  2042 &  2948 &  2866 &  3210 &  3268 &  2457 &  2828 &
3570
  &  3529 &  3819 &  3849 &  3162 &  3467 &  4108 &  4086 &  4335 &
4353 &
 3749\\
 3079 &  3221 &  2190 &  3354 &  3287 &  3583 &  3633 &  2707 &  3121 &
3904
  &  3868 &  4133 &  4160 &  3462 &  3761 &  4401 &  4380 &  4613 &
4630 &
 4034 \\
\hline
\end{tabular}
\end{table}


\end{document}